\documentclass[12pt,a4paper]{article}
\usepackage{mathrsfs}
\usepackage{epsfig}
\usepackage{amstext}
\usepackage{amsmath}
\usepackage{tabularx}
\usepackage{graphics}
\usepackage{graphicx}
\usepackage{float}
\usepackage{multirow}
\usepackage{tabularx}
\DeclareGraphicsExtensions{.eps,.ps,.jpg,.bmp}
\usepackage{color}

\begin{document}
\title{ Pair production of the Elementary Goldstone Higgs boson at the LHC}
\author{Chong-Xing Yue, Lu Ma, Ya-bing Zuo and Yu-chen Guo \\
{\small Department of Physics, Liaoning  Normal University, Dalian
116029, P. R. China}
\thanks{E-mail:cxyue@lnnu.edu.cn}}
\date{\today}

\maketitle
\begin{abstract}
The Elementary Goldstone Higgs (EGH) model is a perturbative extension of the standard model (SM), which identifies the EGH boson as the observed Higgs boson. In this paper, we study pair production of the EGH boson via gluon fusion at the LHC and find that the resonant contribution of the heavy scalar is very small and the SM-like triangle diagram contribution is strongly suppressed. The total production cross section mainly comes from the box diagram contribution and its value can be significantly enhanced with respect to the SM prediction.

\end{abstract}
\newpage

\noindent{\bf 1. Introduction }\vspace{0.5cm}

The discovery of a Higgs boson at the LHC \cite{1}, which in principle seems to be the Higgs boson of the standard model (SM), opens a new window to physics related to the Higgs boson within and beyond the SM. The possibility of detailed and careful study of Higgs physics in the TeV range is higher than past years with the upgrade of the CMS and ATLAS detectors \cite{2}. For the current and future high-energy experiments, such as the LHC Run II and $e^{+}e^{-}$ colliders, one of their main goals is to measure the processes which give information on the Higgs boson couplings to fermions, gauge bosons and its self-coupling and to compare with the most accurate SM predictions. Any deviation from the SM predictions could be valuable information about physics beyond the SM.

To address the SM gauge hierarchy problem, several alternative paradigms have been put forward. The elementary Goldstone Higgs (EGH) scenario \cite{3,4} is one possible paradigm, which is based on an elementary scalar sector with a global symmetry larger than that in the SM. This symmetry is explicitly broken by the couplings with the electroweak (EW) gauge currents and SM Yukawa interactions. Under radiative corrections, this symmetry breaking will align the vacuum with respect to the EW symmetry. The observed Higgs boson is a fundamental pseudo Nambu-Goldstone (PNG) boson of the global symmetry breaking. It can obtain a light mass through radiative corrections which could cause the symmetry breaking  and explain the origin of the known fermion masses.

In the EGH model \cite{3,4}, the elementary Higgs sector of the SM is enhanced to an $SU(4)$ symmetry that breaks spontaneously to $Sp(4)$. The embedding of the EW gauge sector is parameterized by an angle $\theta$. Its value can be fixed by minimizing the quantum corrected effective potential of this model in the presence of the EW and top corrections, which has been dynamically determined to be centered around $\theta\approx0.018$ \cite{4}.

The EGH model is a perturbative extension of the SM, in which the Higgs boson is a fundamental particle, like in the SM, but the mechanism of symmetry breaking is completely different. Thus, it might produce rich new phenomenology at the LHC Run II or future high energy collider experiments, which has not been fully studied in the literature. The relations of the EGH idea with unification scenario, the relaxation leptogenesis mechanism, and supersymmetry have been studied in references \cite{5,6} and \cite{7}, respectively. SM Higgs inflation has also been discussed in the context of the EGH model~\cite{8}.

Pair production of the Higgs boson is well known for its sensitivity to the tri-Higgs coupling, providing a way to test the structure of Higgs potential and further EW symmetry breaking mechanism. In the SM, Di-Higgs production at the LHC, mainly from gluon fusion, arises from both triangle and box loop contributions, which interfere destructively, causing a suppression of the total production rate from the naive estimate \cite{9.1,9.2}. Thus, sizable production of Di-Higgs directly implies a new physics signature \cite{10}. The main goal of this paper is to examine pair production of the EGH boson at the LHC. We find that, although its trilinear coupling with respect to the SM one is strongly suppressed, the box loop contributions, the resonant contributions of the heavy scalar and the constructive interference effects can make up this suppression and enhance the production rate of EGH boson to a sizeable level.

This paper is structured as follows: in Section 2, we summarize the main features of the EGH model and show the relevant couplings. The possible decay channels of the EGH boson $H_1$, which is identified as the observed Higgs boson, and the heavy scalar $H_2$ are discussed in Section 3. Section 4 studies pair production of the EGH boson at the LHC and our conclusions are given in section 5.

\vspace{0.5cm} \noindent{\bf 2. Main features of the EGH model  }

\vspace{0.5cm}For the EGH model \cite{3,4}, the Higgs sector is embedded into a $SU(4)\rightarrow Sp(4)$ pattern of chiral symmetry breaking. The SM Higgs doublet is part of the $SU(4)/Sp(4)$ coset, while the EW symmetry, $SU(2)_L\times U(1)_Y$, is embedded in $SU(4)$. The SM Higgs boson is identified with one of the Goldstone bosons which acquires mass via a slight vacuum misalignment mechanism induced by quantum corrections. The most general vacuum $E_\theta$ of this model can be expressed as
\begin{eqnarray}
E_\theta=E_B\cos\theta +E_H\sin\theta  =-E_\theta^T,
\end{eqnarray}
where $0\leq \theta\leq \pi/2$ and the two independent vacua $E_B$ and $E_H$ are defined as
\begin{equation}
E_B=
\left(
\begin{array}{cc}
i\sigma_2 & 0 \\
0 & -i\sigma_2 \\
\end{array}
\right),\quad
E_H=
\left(
\begin{array}{cc}
0 & 1 \\
-1 & 0 \\
\end{array}
\right).
\end{equation}
Here $\sigma_2$ is the second Pauli matrix. The Higgs sector of this model strictly depends on the choice of the vacuum $E_\theta$. The alignment angle $\theta$ is completely determined by the radiative corrections and the requirement that the EGH model reproduces the phenomenological success of the SM, which prefers small values of $\theta$.

The above vacuum structure can be realised by introducing the scalar matrix $M$, which is the two-index antisymmetric irrep  $\sim6$ of $SU(4)$
\begin{eqnarray}
M=[\frac{\sigma+i\Theta}{2}+\sqrt{2}(i\Pi_i+ \widetilde{\Pi}_i) X_\theta^i]E_\theta .
\end{eqnarray}
Where $X_\theta^i$ ($i=1,\ldots,5$) are the broken generators associated with the breaking of $SU(4)$ to $Sp(4)$, reported in Appendix A of Ref.\cite{4}.

In order to embed the EW gauge sector of the SM into the $SU(4)$ group, the EGH model gauges the $SU(2)_L\times U(1)_Y$ part of the chiral symmetry group $SU(2)_L\times SU(2)_R\subset SU(4)$. In this case, the scalars are minimally coupled to the EW gauge bosons via the covariant derivative of $M$
\begin{eqnarray}
D_\mu M=\partial_\mu M-i(G_\mu M+MG_\mu^T)
{, \quad \rm with \quad} G_\mu=gW_\mu^iT_L^i+g'B_\mu T_Y.
\end{eqnarray}
Where the $SU(2)_L$ generators are $T_L^i$ ($i=1,2,3$) and the hypercharge generator is $T_Y=T_R^3$. The kinetic and EW gauge interaction Lagrangian of the scalar sector can be written as
\begin{eqnarray}
\mathcal{L}_{gauge}=\frac{1}{2}Tr[D_\mu M^\dagger D^\mu M],
\end{eqnarray}
which explicitly breaks the global $SU(4)$ symmetry.

In the EGH model \cite{3,4}, the EGH boson is one of the two linear combinations of the PNG bosons $\sigma$ and $\Pi_4$ at low energy, that is
\begin{equation}
\left(
\begin{array}{c}
\sigma \\
\Pi_4 \\
\end{array}
\right)
=
\left(
\begin{array}{cc}
\cos\alpha & -\sin\alpha \\
\sin\alpha & \cos\alpha
\end{array}
\right)
\left(
\begin{array}{c}
H_1\\
H_2
\end{array}
\right).
\end{equation}
Where $H_1$ and $H_2$ are mass eigenstates, $\alpha$ is the mixing angle and its value taken as $0<\alpha<\pi/2$. The lightest of $H_1$ and $H_2$ is identified as the observed Higgs boson with mass $m_h=125.09\pm0.24$ GeV \cite{11}.

The renormalizability of the EGH model together with the perturbative corrections determine dynamically the direction of the vacuum $E_\theta$. By investigating the available parameter space of the scalar sector, Ref.\cite{4} has shown that the preferred value of the vacuum alignment angle $\theta$ is $\theta=0.018_{-0.003}^{+0.004}$, corresponding to the $SU(4)$ spontaneous symmetry breaking scale of $f=13.9_{-2.1}^{+2.9}$ TeV via the phenomenological constant $f\sin\theta=\nu=246$ GeV and the mixing angle $\alpha=1.57$. This means that the EGH boson $H_1$ is taken as the observed Higgs boson and is expressed by $h$, like the SM Higgs boson. It is mainly comprised of the PNG boson $\Pi_4$ with a tiny admixture of $\sigma$, while the heavier scalar $H_2$ is mainly made up of the PNG boson $\sigma$, which is taken as $H$.

The first three of the five PNG bosons $\Pi_i$($i=1,\ldots,5$) become the longitudinal components of the EW gauge bosons $W$ and $Z$, while the fourth is used to constitute the EGH boson $h$. Reference \cite{3} has shown that $\Pi_5$ is a stable massive particle and provides a viable dark matter candidate. For $M_{\Pi_5}=M_{DM}\geq m_h$, the EGH model is compatible with the experimental constraints. In our following numerical calculation, we will take $M_{DM}=m_h$.

The SM normalised coupling strength of the scalars predicted by the EGH model can be written as \cite{3,4}.
\begin{eqnarray}
K_{h[H]}^F=\frac{g_{h[H]ff}}{g_{ff}^{SM}}=\sin(\alpha+\theta)[\cos(\alpha+\theta)],
\end{eqnarray}
\begin{eqnarray}
K_{h[H]}^V=\frac{g_{h[H]VV}}{g_{VV}^{SM}}=\sin(\alpha+\theta)[\cos(\alpha+\theta)],
\end{eqnarray}
\begin{eqnarray}
\mu_h=\frac{\lambda_{hhh}}{\lambda_{hhh}^{SM}}=\frac{M_\sigma^2\nu\cos\alpha}{fm_h^2},\quad
\mu_H=\frac{\lambda_{HHH}}{\lambda_{hhh}^{SM}}=\frac{M_\sigma^2\nu\sin\alpha}{fm_h^2},
\end{eqnarray}
\begin{eqnarray}
\mu_{Hh}=\frac{\lambda_{Hhh}}{\lambda_{hhh}^{SM}}=-\frac{M_\sigma^2\nu\sin\alpha}{3fm_h^2},\quad
\mu_{hD}=\frac{\lambda_{h\Pi_5\Pi_5}}{\lambda_{hhh}^{SM}}=\frac{M_\sigma^2\nu\cos\alpha}{3fm_h^2},
\end{eqnarray}
\begin{eqnarray}
\mu_{HD}=\frac{\lambda_{H\Pi_5\Pi_5}}{\lambda_{hhh}^{SM}}=-\frac{M_\sigma^2\nu\sin\alpha}{3fm_h^2}.
\end{eqnarray}
Here $ff$ denote all of the fermion pairs, $VV=WW$ and $ZZ$, $\lambda_{hhh}^{SM}=3m_h^2/\nu$ is the SM trilinear self-coupling constant of the Higgs boson. In the following section, we will use these relations to consider the possible decay channels of the scalar particles $h$ and $H$, and focus our attention on the branching ratios of the heavy scalar $H$.

\vspace{0.5cm} \noindent{\bf 3. Decays of scalars $h$ and $H$}
\vspace{0.5cm}

From the discussions given in Section 2, we can see that, except the decay mode $\Pi_5\Pi_5$, the decay modes of the EGH boson $h$ are same as those of the SM Higgs boson and its partial decay widths are universally shifted from the SM predictions by a factor $\sin^2(\alpha+\theta)$. Since the decay channel $h\rightarrow\Pi_5\Pi_5$ is kinematically prohibited, the values of the branching ratios $BR(h\rightarrow XX)$ are also same as those of the SM Higgs boson.

For a specific production process and decay mode $i\rightarrow h\rightarrow f$, the Higgs signal strength is defined as
\begin{eqnarray}
\mu_i^f=\frac{\sigma_i\cdot BR^f}{(\sigma_i)^{SM}\cdot(BR^f)^{SM}}.
\end{eqnarray}
Here $\sigma_i(i=ggF, VBF, Wh, Zh, tth)$ and $BR^f(f=ZZ, WW, \gamma\gamma, \tau\tau, bb, \mu\mu)$ are respectively the production cross section for $i\rightarrow h$ and the branching ratio for the decay process $h\rightarrow f$. The subscript ``SM" refers to their respective SM predictions. The values of $\mu_i^f$ are the same for all production processes $i$ and decay channels $f$ in the EGH model, and $\mu_i^f=\mu=\sin^2(\alpha+\theta)$. A fit to the combined ATLAS and CMS data at the center-of-mass (c.m.) energies $\sqrt{s}=7$ and 8 TeV give the best fit value of the Higgs signal strength $\mu$ as $\mu=1.09_{-0.1}^{+0.11}$ \cite{12,13}. The preferred values $\theta=0.018_{-0.003}^{+0.004}$ and $\alpha=1.57$ given by Ref.\cite{4} satisfy this experimental constraint.

The heavy scalar $H$ can decay to the SM gauge bosons and fermions with partial widths of
\begin{eqnarray}
\Gamma(H\rightarrow XX)=\cos^2(\alpha+\theta)\Gamma^{SM}(H\rightarrow XX),
\end{eqnarray}
where $\Gamma^{SM}(H\rightarrow XX)$ is the total decay width of the SM Higgs boson into the $XX$ final states evaluated at the $H$ mass $M_H$. At tree level, the partial decay widths for the processes $H\rightarrow hh$ and $\Pi_5\Pi_5$ can be written as
\begin{eqnarray}
\Gamma(H\rightarrow hh)=
\Gamma(H\rightarrow \Pi_5\Pi_5)=\frac{M_\sigma^4\sin^2\alpha}{32\pi f^2M_H}\sqrt{1-\frac{4m_h^2}{M_H^2}}.
\end{eqnarray}
\begin{figure}[hb!]
\centering
    \includegraphics[width=0.6\textwidth]{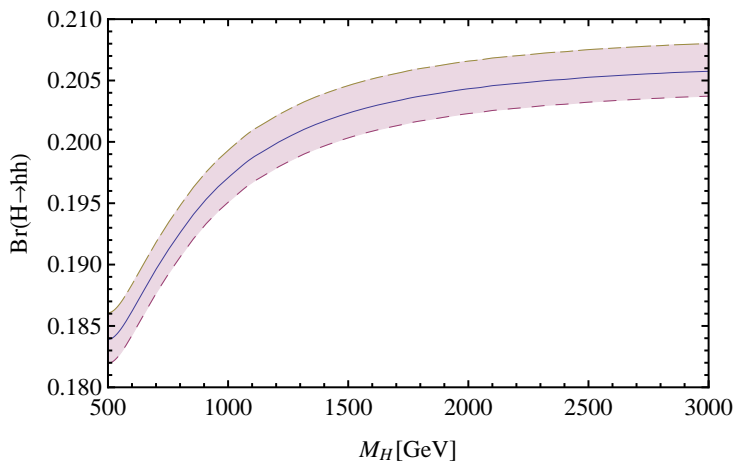}
\caption{The branching ratio $BR(H\rightarrow hh)$ as a function of the mass $M_H$ for $\alpha=1.57$ and $\theta=0.018$ within the statistical error on $\theta$ (the pink region).}
\label{fig:1}
\end{figure}
In the above equation we have taken $M_{\Pi_5}\approx m_h$. Then the total decay width of the heavy scalar $H$ is given by
\begin{eqnarray}
\Gamma_H=\cos^2(\alpha+\theta)\Gamma^{SM}(M_H)+2\Gamma(H\rightarrow hh).
\end{eqnarray}

Considering the heavy scalar $H$ mainly from the PNG boson $\sigma$, we take $M_H\approx M_\sigma$ and assume its values in the range of 1 TeV $\sim$ 3 TeV, using Madgraph5/aMC@NLO program \cite{14} to give our numerical results. In Fig. \ref{fig:1} we plot the branching ratio $BR(H\rightarrow hh)$ as a function of the mass parameter $M_H$ for $\alpha=1.57$ and $\theta=0.018_{-0.003}^{+0.004}$, where the breaking scale $f$ is taken as $f=\nu/\sin\theta$. In most of the parameter space, the decay channel $H\rightarrow hh$ is one of the dominant ones of the heavy Higgs boson and its branching ratio value is about 20\%.
\begin{figure}[ht!]
    \centering
    \includegraphics[width=0.6\textwidth]{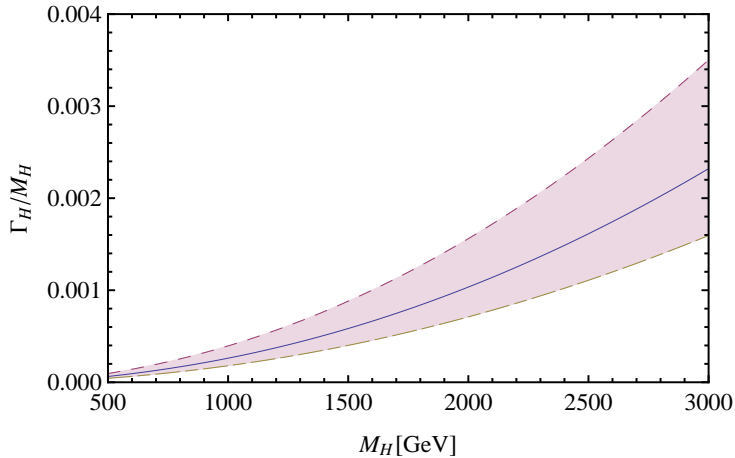}
    \caption{The ratio $\Gamma_H/M_H$ as a function of $M_H$ for $\alpha=1.57$ and $\theta=0.018_{-0.003}^{+0.004}$.}
    \label{fig:2}
\end{figure}

For $\alpha=1.57$ and $\theta=0.018_{-0.003}^{+0.004}$, the allowed range of the ratio $\Gamma_H/M_H$ is shown as a function of $M_H$ in Fig. \ref{fig:2}. In almost all the parameter space, the value of $\Gamma_H/M_H$ is smaller than 0.4\%. Thus, we can safely use the narrow width approximation method to calculate the production cross section of the process $ gg\rightarrow H\rightarrow hh$.

\vspace{0.5cm} \noindent{\bf 4. Pair production of the EGH boson}
\vspace{0.5cm}

Pair production of the EGH boson $h$ at the LHC is mainly induced by two sources: one comes from the diagrams depicted in Fig. 3(a) and 3(b), which are similarly to the SM Di-Higgs production, while the other comes from the resonant process through the heavy scalar $H$ decay $H\rightarrow hh$, as shown in Fig. 3(c). The particles in the loops are the heavy SM fermions, such as top and bottom quarks.

\begin{figure}
\includegraphics[width=0.44\textwidth]{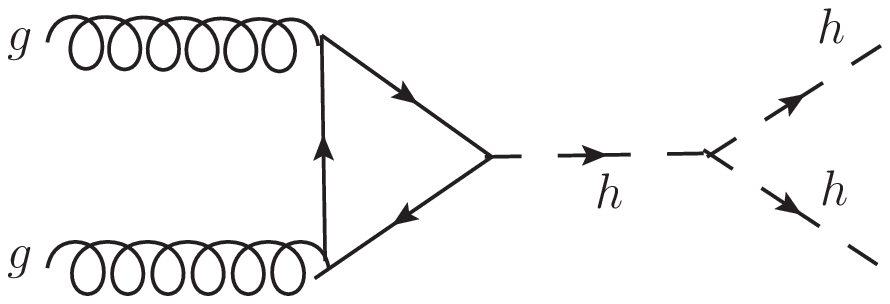}
\includegraphics[width=0.36\textwidth]{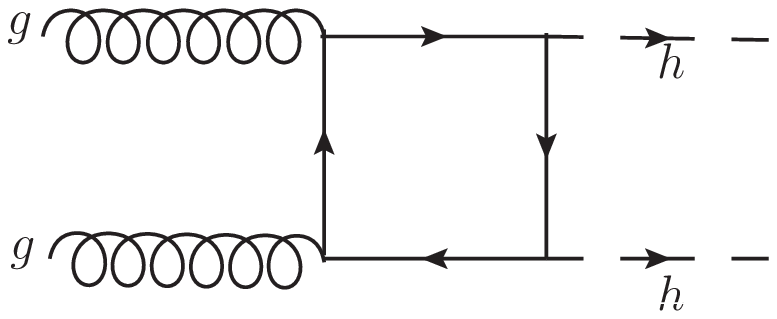}\\
\centering
(a)\hspace{6.5cm}(b) \vspace{0.3cm}\\
\centering
\includegraphics[width=0.44\textwidth]{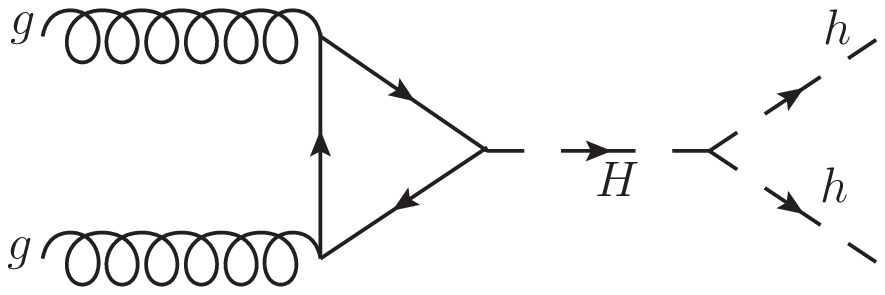}\\
\centering (c)
\caption{Feynman diagrams for Higgs bosons pair production in the EGH model.} \label{fig:3}
\end{figure}

In the EGH model, the trilinear coupling of the EGH boson and its couplings to the SM fermions are modified from the relevant SM couplings. Thus, compared with the SM prediction, the production cross section induced by Fig. 3(a) and 3(b) is changed as
\begin{eqnarray}
\sigma_1=\sigma_T(K_h^F)^2\mu_h^2+\sigma_B(K_h^F)^4+2\cos\alpha_I\sqrt{\sigma_T\sigma_B}(K_h^F)^3\mu_h.
\end{eqnarray}
 Where $\sigma_T$, $\sigma_B$ and $\alpha_I$ respectively represent the cross sections only from the triangle and box graphs, and the interference angle $\alpha_I$. There is $K_h^F= \mu_h=1$ for the SM. In the SM, Ref.\cite{14} has calculated the values of $\sigma_T$  and $\sigma_B$ for various c.m. energy $\sqrt{s}$ and shown that $\cos\alpha_I$ is almost independent of the c.m. energy $\sqrt{s}$, the scale and the parton distribution functions (PDFs), and $\cos\alpha_I=-0.898$ for $\sqrt{s}$=13 TeV.

 The second source contributing to EGH pair production at the LHC is resonant contribution from Fig. 3(c). If a new scalar is sufficiently heavy and can decay on-shell into two SM-like Higgs bosons, it has been shown that the new scalar can significantly enhance the Di-Higgs production rate over the SM prediction via its resonant production by gluon-gluon fusion \cite{15,16,17}. Using the narrow width approximation, the resonant production cross section can be approximately written as
 \begin{eqnarray}
\sigma_2=\sigma (pp\rightarrow H\rightarrow hh)=
\sigma (gg\rightarrow h)_{m_h\rightarrow M_H}\times(K_H^F)^2\times BR(H\rightarrow hh).
\end{eqnarray}
Where $\sigma (gg\rightarrow h)$  is the cross section of single production for the SM Higgs boson via the gluon-gluon fusion process evaluated at the $H$ mass $M_H$.

\begin{figure}[htb!]
\centering
    \includegraphics[width=0.6\textwidth]{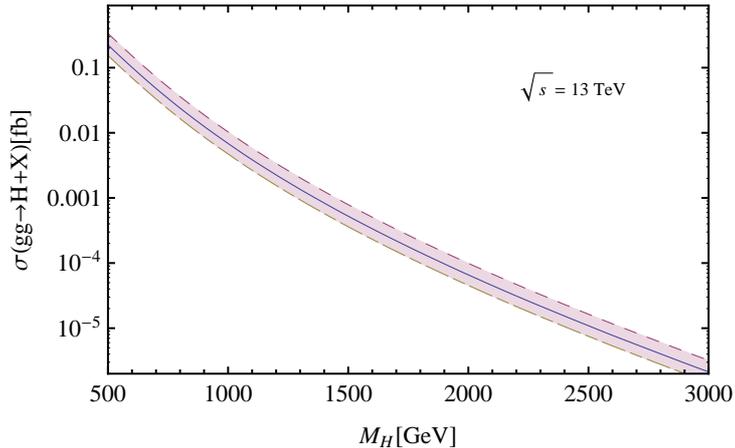}
\caption{The cross section $\sigma_2$ contributed only by the heavy scalar $H$ as a function of $M_H$ for  $\alpha=1.57$ and $\theta=0.018_{-0.003}^{+0.004}$.}
\label{fig:4}
\end{figure}
The interference effects between the two kinds of contributions, the SM-like and heavy scalar contributions, might be significantly large \cite{17}. So in our numerical calculation, we will include these effects. Then the total cross section of pair production of the EGH boson at the LHC is written as
\begin{eqnarray}
\sigma^{EGH}=\sigma_1+\sigma_2+\sigma_{12},
\end{eqnarray}
where $\sigma_{12}$ represents the cross section from the interference contributions. Using the PDFs of CT14 \cite{19} and the NNLO $\kappa$-factor as $\kappa$=2.30 for $\sqrt{s}$=13 TeV \cite{20}, we plot in Fig. 4 the cross section $\sigma_2$ contributed by the heavy scalar $H$ resonant contributions as a function of the mass parameter $M_H$ for  $\alpha=1.57$ and $\theta=0.018_{-0.003}^{+0.004}$. From Fig. \ref{fig:4}, the production cross section is very small and $\sigma_2\leq$0.01 $fb$ for $M_H\geq$1 TeV, which cannot be detected at the LHC in the near future. This is unlike the resonant enhancement arising from a new scalar, which is because of the suppression factor $\cos^2(\alpha+\theta)$.

\begin{figure}[htb!]
\centering
    \includegraphics[width=0.6\textwidth]{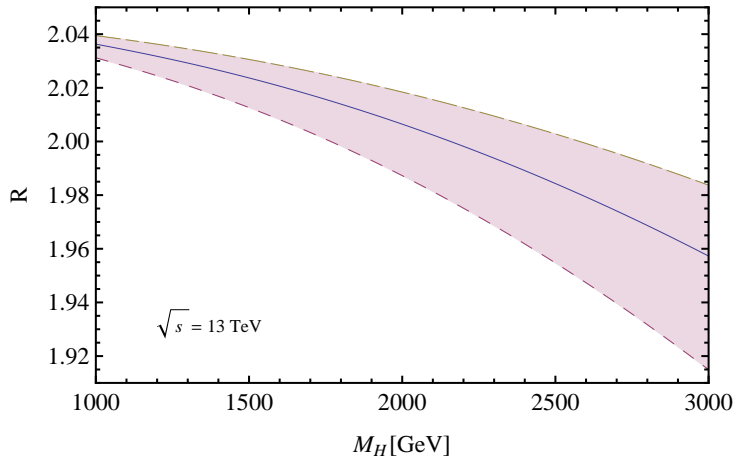}
    \caption{The ratio $R=\sigma^{EGH}/\sigma^{SM}$ as a function of $M_H$ for $\alpha=1.57$ and $\theta=0.018_{-0.003}^{+0.004}$.}
    \label{fig:5}
\end{figure}

In the EGH model, the trilinear self-coupling of the SM-like Higgs boson with respect to the SM one is strongly suppressed, while its couplings with the SM fermions are almost the same as those for the SM Higgs boson. Thus, the cross section of pair production of the EGH boson at the LHC is dominated by the box contributions [Fig. 3(b)] and its destructive interference with the triangle diagram [Fig. 3(a)] is suppressed. Furthermore, the interference contribution between Fig. 3(b) and Fig. 3(c) is positive. So, in the context of the EGH model, the production cross section is larger than that in the SM. Our numerical results show that it is indeed this case. Fig. \ref{fig:5} shows the ratio $R=\sigma^{EGH}/\sigma^{SM}$ as a function of the mass parameter $M_H$, where the value of the SM Di-Higgs production cross section $\sigma^{SM}$ is taken as 37.91 $fb$ calculated at NNLL level for $m_h$=125.09 GeV and the c.m. energy $\sqrt{s}$=13 TeV \cite{20}. One can see from this figure that the production rate for pair production of the SM-like Higgs boson in the EGH model is significantly larger than the SM prediction. For $\alpha=1.57$, $\theta=0.018_{-0.003}^{+0.004}$ and 1 TeV $\leq M_H\leq$ 3 TeV, the value of the ratio $R=\sigma^{EGH}/\sigma^{SM}$ is in the range of 2.04 $\sim$ 1.91.

The triangle and box diagrams contributing to Di-Higgs production have different phase space dependencies and distinct kinematic features. The triangle contribution would peak around $\sqrt{\hat{s}}=m_h$ and become subleading at larger invariant masses. A modified triple Higgs self-coupling of the SM-like Higgs boson will mostly reveal itself at low invariant masses, while shifted Yukawa couplings will typically become more apparent in the larger invariant mass region. The small change of the triple Higgs self-coupling will mainly affect the low invariant mass region which will be removed by the cuts. However, when the triple Higgs self-coupling is nearly turned off, the high invariant mass region signal event number will not increase as much as the total cross section \cite{ATL-2017-001}. In the EGH model, the trilinear self-coupling of the SM-like Higgs boson is strongly suppressed, while its couplings with the SM fermions are almost the same as those for the SM Higgs boson. Furthermore, the production cross section contributed by the heavy scalar $H$ is very small for $M_H\geq$1 TeV. Thus, in the context of the EGH model, the production rates of the final states, such as $ b \overline{b} \gamma\gamma$ and $ b \overline{b} \tau\tau$, can be enhanced with respect to the SM ones. As a reasonable estimation, the value of the ratio $R$ for the signal event number is in the range of 1.6 $\sim$ 1.7 after the cuts effect is considered.

Currently, the ATLAS and CMS collaborations have been searching for the signal of pair production of the Higgs boson at the LHC and also focusing on resonance-enhanced production mechanisms \cite{21}. Upper limits on the production cross sections for some categories of signal final states have been obtained. In the near future, the LHC will give more meaningful data, which can be used to test the SM and further probe new physics beyond the SM. Our work will help to examine the EGH model via pair production of the SM-like Higgs boson at the LHC.

\vspace{0.5cm} \noindent{\bf 5. Conclusions  }

\vspace{0.5cm}The EGH model is a perturbative extension of the SM featuring an EGH boson and dark matter particle, which is theoretically well-motivated and phenomenologically viable. In this model, the EGH boson is taken as the observed  Higgs boson, which has almost the same couplings as the SM fermions and the EW gauge bosons as those for the SM Higgs boson. Thus, this model can easily satisfy the experimental constraints from the Higgs signal data at the LHC.

The trilinear self-coupling of the EGH boson is suppressed with respect to the SM one, which modifies the cross section for pair production of the EGH boson from the SM prediction at the LHC. In addition, the existence of the heavy scalar $H$ in the EGH model gives an additional contribution to this cross section. So, in this paper , we have studied pair production of the EGH boson via the gluon-gluon fusion process at the LHC.

Our numerical results show that, since the couplings of the heavy scalar $H$ to the SM fermion pairs are strongly suppressed by the factor $\cos(\alpha+\theta)$ for $\alpha=1.57$ and $\theta=0.018_{-0.003}^{+0.004}$, its resonant contribution to the Di-Higgs production cross section is negligible and its value is smaller than 0.1 $fb$ in most of the parameter space, while its interference contribution with the box diagram is positive and cannot be neglected. In the EGH model, the total production cross section of the SM-like Higgs boson pair is larger than the SM prediction. The value of the ratio $R=\sigma^{EGH}/\sigma^{SM}$ is in the range of 2.04$\sim$1.91 for $\alpha=1.57$, $\theta=0.018_{-0.003}^{+0.004}$ and 1 TeV $\leq M_H\leq$ 3 TeV. The EGH model might be probed or ruled out by the LHC via the Di-Higgs production process in the near future.

\section*{Acknowledgments} \hspace{5mm}This work was
supported in part by the National Natural Science Foundation of
China under Grants No.11275088 and 11545012, and  the Natural Science Foundation of the Liaoning Scientific Committee
(No. 2014020151).

\end{document}